\documentclass[
nofootinbib,
twocolumn
]{revtex4-2}
\usepackage{amsmath,amssymb,amsfonts,graphicx}
\usepackage[caption=false]{subfig}
\usepackage{ mathrsfs }
\usepackage{dsfont}
\usepackage{algorithm}
\usepackage{mathdots}
\usepackage{algpseudocode}
\usepackage{xr-hyper}
\usepackage[bottom]{footmisc}
\usepackage{xcolor}
\definecolor{darkred}{rgb}{0.801, 0.0, 0.0}
\usepackage[colorlinks=true,breaklinks=true, urlcolor=darkred,citecolor=darkred,anchorcolor=darkred,linkcolor=darkred]{hyperref}
\usepackage{breakcites}
\usepackage{ulem}
\def\eqref#1{Eq.$~$\textcolor{darkred}{ \ref{#1}}}

\def\vphi{\varphi}

\begin{document}

\title{Exact and approximate fluxonium array modes}

\author{Stephen Sorokanich}
\affiliation{Applied and Computational Mathematics Division, National Institute of Standards and Technology, Gaithersburg, MD 20899-8910, USA}
\email{stephen.sorokanich@nist.gov}

\author{Max Hays}
\affiliation{Research Laboratory of Electronics, Massachusetts Institute of Technology, Cambridge, MA 02139, USA}
\email{maxhays@mit.edu}

\author{Neill~C.~Warrington}
\affiliation{Center for Theoretical Physics, Massachusetts Institute of Technology, Cambridge, MA 02139, USA}
\email{ncwarrin@mit.edu}

\date{\today }

\begin{abstract}
    We present an exact solution for the linearized junction array modes of the superconducting qubit fluxonium in the absence of array disorder. This solution holds for arrays of any length and ground capacitance, and for both differential and grounded devices.  Array mode energies are determined by roots of convex combinations of Chebyshev polynomials, and their spatial profiles are plane waves. We also provide a simple, approximate solution, which estimates array mode properties over a wide range of circuit parameters, and an accompanying Mathematica file that implements both the exact and approximate solutions. 
\end{abstract}
\maketitle

\section{Introduction}\label{Intro}

The Josephson junction is arguably the most important building block of modern superconducting circuits. 
Large arrays of junctions have found applications in quantum information processing, quantum metrology, quantum-limited amplification, and many-body simulation~\cite{krantz2019quantum,crescini2023evidence,CRPHYS_2016__17_7_740_0,mehta2023down}. 
In superconducting quantum computing, serially-connected arrays of junctions are used to create so-called ``superinductors'', reactive elements with impedances greater than the resistance quantum~\cite{manucharyanThesis,masluk2012microwave}. 
Along with single junctions and capacitors, superinductors compose the contemporary toolbox of circuit elements from which superconducting qubits are constructed. 
In particular, superinductors are critical components of certain species of qubits such as the fluxonium, $\cos 2\phi$, and zero-pi~\cite{manucharyan2009fluxonium,smith2020superconducting,gyenis2021experimental}.

While it is often sufficient to treat a junction-array-based superconductor as a single linear inductor, it is important to remember that the phase drop across each junction in the array is fundamentally a separate degree of freedom. 
Noise coupled to these degrees of freedom can lead to decoherence of a qubit encoded in the circuit, for instance via the Aharonov-Casher effect~\cite{manucharyan2012evidence,PhysRevB.102.014512,di2021efficient, randeria2024dephasing}. 
Moreover, these internal degrees of freedom of the superinductor are strongly coupled, forming higher-order ``array modes'' of the circuit. 
When designing qubit circuits, one typically  wishes to keep the frequency of these array modes much higher than the system temperature, such that they can be safely neglected when considering the circuit dynamics. 
However, accurately predicting the array mode frequencies requires solving the full Hamiltonian of the circuit, including the degrees of freedom associated with the many junctions of the superinductor. 

In this work we focus on the array modes of fluxonium, a superconducting qubit formed by shunting a single junction with a superinductor and a capacitor~\cite{Manucharyan_2009} (see Fig.~\ref{fig:simp}). 
When constructing the superinductor from serially-conncted junctions, it is typical to choose device parameters such that a single dominant mode, the ``superinductance mode", corresponding to a correlated fluctuation of phases across the entire array, governs the low-energy behavior, while all other modes, the so-called ``array modes", have higher energies and are thus comparatively suppressed. The low-energy Hamiltonian for the superinductance mode is simple compared to the full Hamiltonian and is given by
\begin{equation}
\label{eq:simp-ham}
    H = - 4 E_C \partial_{\vphi}^2 - E_J \text{cos}(\vphi - \vphi_{\text{ext}})+ \frac{E_L}{2} \vphi^2~,
\end{equation}
where $E_C, E_J, E_L$ are the effective capacitive, Josephson, and inductive energies and $\vphi_{\text{ext}}$ is related to the external flux threading the loop through $\vphi_{\text{ext}} = 2\pi \Phi_{\text{ext}}/\Phi_0$, where $\Phi_0 = h/2e$ is the superconducting flux quantum. 
While this Hamiltonian has been used to describe the dynamics of fluxonium devices with great success, \cite{Manucharyan_2009,PhysRevX.9.041041,zhang2021universal},  
for a more complete representation of the device physics one must treat the phase drop across each array junction as a separate degree of freedom~\cite{masluk2012microwave, symmetries-and-collective, Viola_2015, di2021efficient, hazard2019nanowire}. 
These considerations become especially important as $E_L$ shrinks (inductance grows) because the array modes move to lower energy and thus become progressively more relevant to the circuit dynamics. 

Here we present an analytic solution to the array modes of the fluxonium model schematically depicted in Fig.~\ref{fig:comp}. 
Previous theoretical works have used various approximations to examine the structure and noise properties of fluxonium array modes~\cite{masluk2012microwave, symmetries-and-collective, Viola_2015, di2021efficient, hazard2019nanowire}. 
Building off these works, we present an analytic solution to array modes of the full fluxonium Hamiltonian as well as a useful approximation scheme. In Section \ref{Fluxonium} we define the model, tracking the effect of array modes on the quantum Hamiltonian. In Section \ref{esys} we derive the exact array modes of a differential fluxonium device in the presence of ground capacitances, then show how the array modes of a grounded fluxonium device is contained in this result. In Section \ref{approx} we present a simple approximation scheme for array modes. In Section \ref{conclusion} we present our conclusions.

\section{Fluxonium}\label{Fluxonium}


\begin{figure}[t]
\begin{center}
	\subfloat[\label{fig:simp}]{
            \includegraphics[width=0.3\columnwidth]{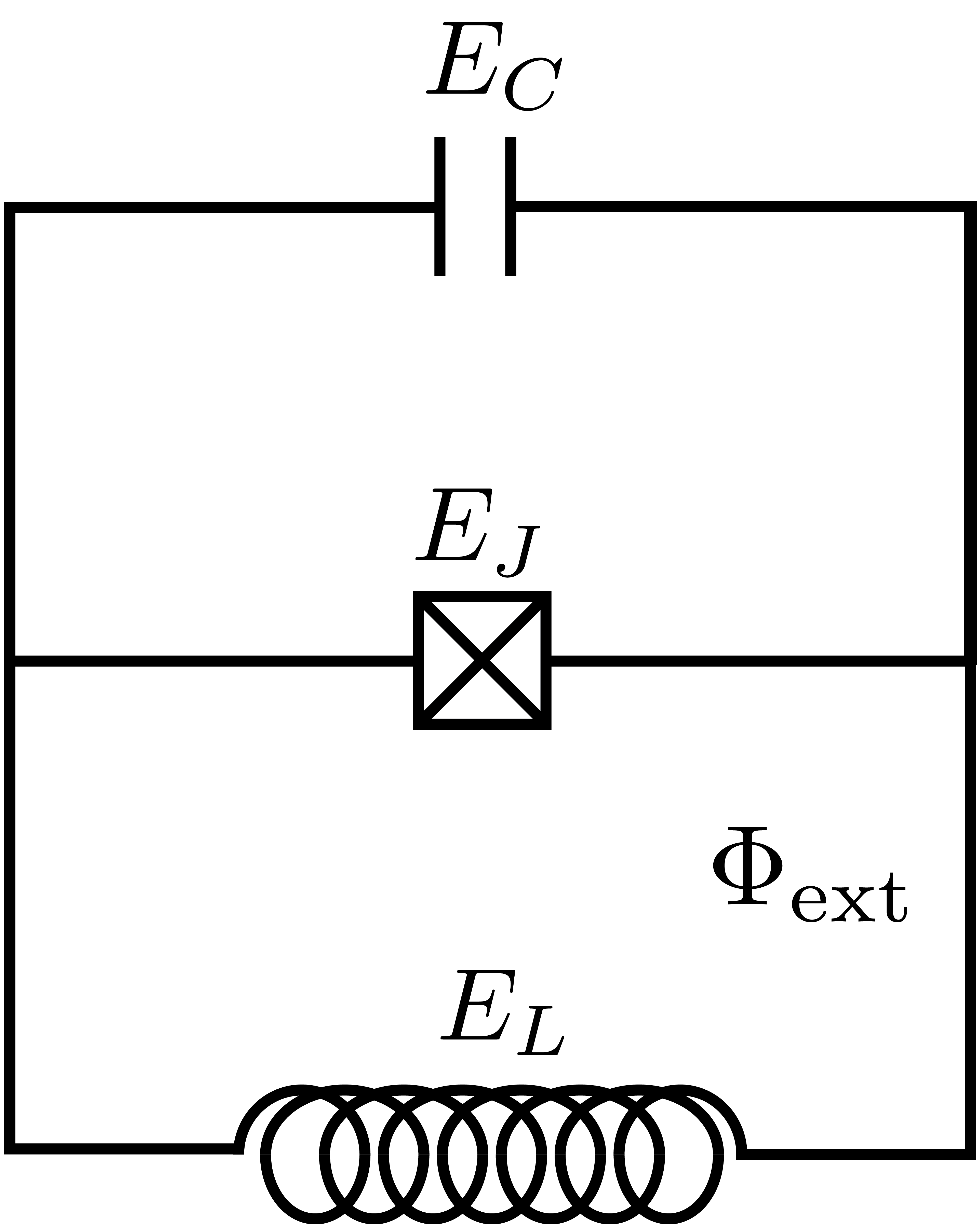}
            }
	\subfloat[\label{fig:comp}]{
            \includegraphics[width=0.65\columnwidth]{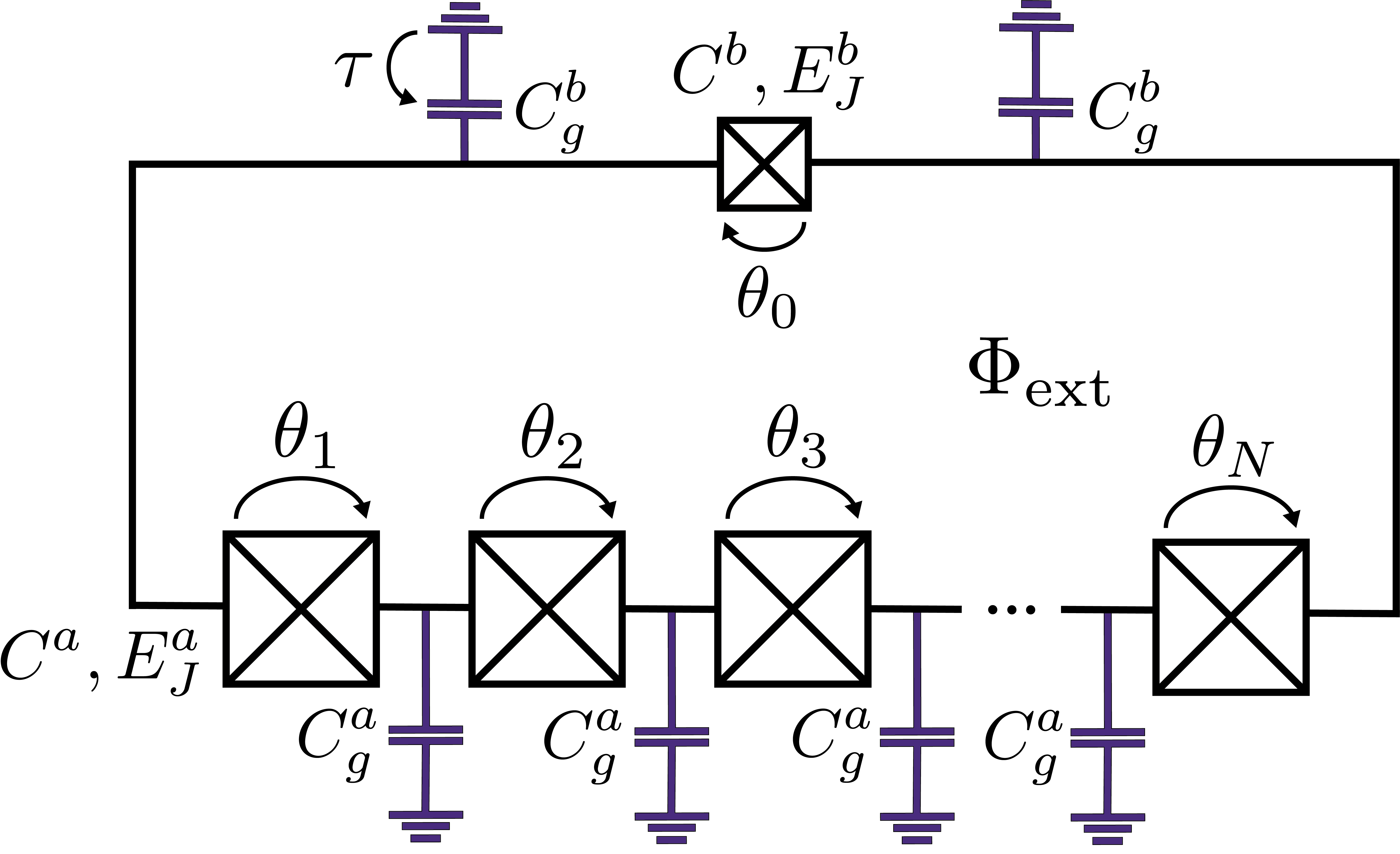}
            }
\end{center}
    \caption{\protect\subref{fig:simp} Simplified fluxonium circuit diagram 
    \protect\subref{fig:comp} Full fluxonium circuit diagram. Array junctions are assumed to have identical capacitances and Josephson energies, $C^a, E_J^a$, while the smaller junction has different values $C^b, E_J^b$ (to simplify the diagram we include any capacitance shunting the small junction in $C^b$). Similarly, the array junctions are supposed to couple to ground through capacitances $C_g^a$ while the smaller junction couples to ground capacitance $C_g^b$ (shown in purple). An external flux $\Phi_{\text{ext}}$ threads the loop. 
    }
    \label{fig:fluxonium}
\end{figure}

A diagram of fluxonium is shown in Fig. \ref{fig:fluxonium}: it consists of a single superconducting loop containing $N+1$ junctions threaded by an external flux $\Phi_{\text{ext}}$. 
Indicated in the figure, $\theta_0$ represents the gauge invariant branch phase across the small junction while $\theta_1, \dots, \theta_N$ represent phases across the $N$ array junctions. The phases are constrained by the fluxoid quantization condition to satisfy $\sum_{m=0}^N \theta_m + \vphi_{\text{ext}} = 2 \pi z, ~z\in \mathbb{Z}$, and we employ this condition from the beginning to eliminate $\theta_0$ from all expressions. 

In this paper we consider differential, or floating, fluxonium, where the overall voltage may fluctuate in time.  This is reflected by the presence of the $\tau$ variable in Fig. \ref{fig:fluxonium}, which is a reference node variable such that $\dot \tau \neq 0$. Grounded fluxonium devices, on the other hand, have $\dot \tau = 0$. Both grounded and differential devices are being built today \cite{PhysRevX.13.031035,PhysRevX.9.041041}, and interestingly the array mode physics of a grounded fluxonium circuit is contained in that of differential one in the absence of array disorder. While our analysis focuses on differential fluxonium, we show at the end of the next section how the grounded can be obtained from the differential case. 

We denote the Lagrangian of a differential fluxonium device by
\begin{equation}
\label{eq:lagrangian}
    L = T- U~,
\end{equation}
where $T, \, U$ are the kinetic and potential energies, respectively. We consider a potential energy
\begin{equation}
\label{eq:potential}
    U = 
    -E_J^a \sum_{m=1}^N \text{cos}\big( \theta_m\big) 
    - E_J^b \text{cos}\big(\sum_{m=1}^N \theta_m - \vphi_{\text{ext}} \big)~,
\end{equation}
and kinetic energy $T = T_0 + T_g$, where 
\begin{align}
\label{eq:kinetic}
T_0 & = \frac{1}{2} \Big( \frac{\hbar}{2e} \Big)^2 \Bigg[ C^a \sum_{i=1}^N \dot{\theta}_i^2  + C^b \left(\sum_{i=1}^N \dot{\theta}_i \right)^2 \Bigg]~ \nonumber \\
T_g & = \frac{1}{2} \Big( \frac{\hbar}{2e} \Big)^2 \Bigg[ C_g^a \sum_{i=1}^{N-1}  \Big(\dot{\tau} + \sum_{j=1}^{i}\dot{\theta}_j \Big)^2  \nonumber \\
& ~~~~~~~~~~~~~ + C_g^b \dot{\tau}^2 + C_g^b \Big(\dot{\tau} + \sum_{j=1}^{N}\dot{\theta}_j \Big)^2 \Bigg]~.
\end{align}
This system has been previously considered 
\cite{symmetries-and-collective,Viola_2015}. $T_0$ represents the kinetic energy of the junctions themselves while $T_g$ models capacitive coupling to ground \cite{PhysRevB.102.014512,flux-qubit-revisited,devoret_quant_fluct,CALDEIRA1983374}. The terms in $T_g$ are represented by the purple capacitors in Fig. \ref{fig:fluxonium}. The first term of $T_g$, proportional to $C^a_g$, accounts for ground capactiances along the array and is represented the purple capacitors at the bottom of the figure. The two remaining terms, proportional to $C_g^b$, account for ground couplings of the small junction and are represented by the purple capacitors at the top of the figure.  

We emphasize that we are not considering the most general model of fluxonium: we assume zero array disorder. We make this assumption in order to obtain an analytic solution for array modes. A more realistic model incorporating array disorder was solved previously \cite{di2021efficient}. This solution is complementary to ours, providing a semi-analytic solution to a more general case. 

Our aim is to provide exact expressions for the array modes of the model of fluxonium model defined by Eqs. \ref{eq:lagrangian}-\ref{eq:kinetic}. These arise in the course of ``circuit quantization", a general method for formulating quantum theories of circuits \cite{devoret_quant_fluct,PhysRevB.69.064503}. The method proceeds in two steps: first, the classical Lagrangian is  transformed into a classical Hamiltonian through a Legendre transformation, and second, the theory is quantized by promoting all dynamical variables appearing in the Hamiltonian to operators on a Hilbert space, and by imposing on conjugate variables the canonical commutation relations. 

The first step, Legendre transforming from Lagrangian to Hamiltonian, is complicated by the coupling between phase variables in Eq. \ref{eq:kinetic}. To accomplish this transformation, we proceed as follows\footnote{Here and throughout we drop unimportant constant terms.}. We first re-write the kinetic energy as
\begin{equation}
\label{eq:compact-kinetic}
T = \frac{1}{2} \Big( \frac{\hbar}{2e}\Big)^2 ~ \dot x^T \mathds{M} \dot x  ~,
\end{equation}
where $x := (\tau, \theta_1,\dots,\theta_N)$. The $(N+1)\times(N+1)$ matrix $\mathds{M}$ is equal to
\begin{equation}
    \mathds{M} = 
    \begin{bmatrix}
        a & b^T \\
        b & C
    \end{bmatrix}~,
\end{equation}
where $a = 2 C^b_g + (N-1) C^a_g$ is a number, $b$ is a vector with components $b_m = C^b_g +C^a_g(N-m)$, $(m=1,\dots,N)$
and $C = C_1 + C_2 + C_3$ is an $N\times N$ matrix. Here, $\big(C_1\big)_{ij} = C^a \delta_{ij}$, $\big( C_2 \big)_{ij} = C^b + C^b_g$, and
\begin{align*}
    C_3 & = C^a_g 
    \begin{bmatrix}
    N-1&N-2&\dots&1&0\\
    N-2&N-2&\dots&1&0 \\
    \vdots&\vdots&\ddots&\vdots\\
    1&1&\dots&1&0\\
    0&0&\dots&0&0\\ 
    \end{bmatrix}~.
\end{align*}
Since the potential energy Eq. \ref{eq:potential} is independent of $\tau$, its conjugate momentum is conserved, and $\tau$ can be eliminated from the Lagrangian entirely. Defining the total charge 
\begin{equation}
    n_{\text{tot}} = \frac{\partial L}{\partial \dot\tau} = 0~,
\end{equation}
one finds $\dot \tau = - a^{-1} b^T \dot \theta$, where $\theta := (\theta_1,\dots,\theta_N)$. Using this relation to eliminate $\dot \tau$ from the Lagrangian, one obtains
\begin{equation}
\label{eq:new-T}
    L =  
    \frac{1}{2} \Big( \frac{\hbar}{2e}\Big)^2 ~ \dot \theta^T \mathds{C} \dot \theta  - U~, 
\end{equation}
where $\mathds{C} = C + C_4$, $C_4 = - b b^T/a$. We have checked that $\mathds{C}$ is the same capacitance matrix as in \cite{symmetries-and-collective} (using their Eq. D3). Defining conjugate momenta $p_i = \partial L/\partial \dot \theta_i$ and performing the Legendre transformation, one finally obtains the classical Hamiltonian
\begin{equation}
\label{eq:ham}
    H = \frac{1}{2} \Big( \frac{2e}{\hbar}\Big)^2 ~ p^T \mathds{C}^{-1} p + U~.
\end{equation}





Having computed the classical Hamiltonian, one may quantize the theory. What results is a challenging quantum many-body problem which we do not attempt to solve (many-body calculations do exist, though, for example the tensor network calculations of \cite{di2021efficient}). Instead we focus on exactly computing the array modes and tracking their effect on the quantum theory. 
To accomplish the latter we express all quantites in terms of the eigenvalues \& normalized eigenvectors of $\mathds{C}^{-1}$, which we denote as $\lambda_m$ and $v_m$, and solve for in the following section. 

We begin by defining a diagonal matrix $\Lambda_{mn} = \lambda_m \delta_{mn}$ and an orthogonal matrix $\mathds{V}_{mn} = {({v_m})}_n$, so that $\mathds{C}^{-1} = \mathds{V}^T \Lambda \mathds{V}$ and $\mathds{V}^T \mathds{V}= \mathds{1}$. Defining new canonically conjugate variables $P = \mathds{V}p$ and $\Theta = \mathds{V}\theta$, one has $p^T \mathds{C} p = P^T \Lambda P$ and 
\begin{align}
    \theta_m & = \sum_{a=1}^N \Theta_a \mathds{V}_{am} \nonumber \\
    \sum_{m=1}^N \theta_m
    & =
    \Theta_0 \sum_{b=1}^N \mathds{V}_{0b}+\sum_{\mu=1}^{N-1}\sum_{b=1}^N \Theta_{\mu} \mathds{V}_{\mu b}~.
\end{align}
Promoting conjugate variables to operators satisfying $[\Theta_m , P_n] = i \hbar \delta_{mn}$ and defining
\begin{align}
    \vphi &= \Theta_0 \sum_{b=1}^N \mathds{V}_{0 b} \nonumber \\
    \mathcal{N}^{1/2} & =  \sum_{b=1}^N \mathds{V}_{0 b} \nonumber \\
    \mathcal{N}_{\mu}^{1/2} & =  \sum_{b=1}^N \mathds{V}_{{\mu} b}~,
\end{align}
one obtains the quantum Hamiltonian for fluxonium:
\begin{align}\label{eq:exact-ham}
    \hat H = & 
    - 4 E_C {\partial^2_{\vphi}} - 4 \sum_{\mu=1}^{N-1} E_{\mu} \partial_{\mu}^2 \nonumber \\ 
    & - E^a_J \sum_{m=1}^N \text{cos}\Big( \vphi \frac{\mathds{V}_{0m}}{\mathcal{N}^{1/2}} + \sum_{\mu=1}^{N-1} \Theta_{\mu} \mathds{V}_{\mu m} \Big) \nonumber \\
    & - E^b_J \text{cos}\Big( \vphi - \vphi_{\text{ext}} + \sum_{\mu=1}^{N-1} \Theta_{\mu} \mathcal{N}^{1/2}_{\mu} \Big) ~,
\end{align}
where $E_C = \mathcal{N} e^2 \lambda_0 / 2$, $E_{\mu} = e^2 \lambda_{\mu} / 2$, and $\partial_{\mu} = \partial/\partial \Theta_{\mu} $. Similar expressions for the fluxonium Hamiltonain were found previously \cite{di2021efficient}. 

At this point no approximations have been made. When ground capacitances are zero, however, the eigensystem of $\mathds{C}^{-1}$ simplifies. One finds that $\lambda_0 = 1/(C^a + N C^b)$ and $\lambda_{\mu} = 1/C^a$ for every $\mu$; the eigenvector for the superinductance mode is constant, with components ${(v_0)}_m = \mathds{V}_{0m} = N^{-1/2}$, and within the degenerate subspace a basis of array modes can be chosen so that ${(v_{\mu})}_m = \mathds{V}_{{\mu}m} = (2/N)^{1/2}\text{cos}(\mu \pi / N (m-1/2))$ \cite{symmetries-and-collective}. Thus $\mathcal{N} = N$, and $\mathcal{N}_{\mu} = 0$. This choice of array mode basis eliminates them from the final line of \eqref{eq:exact-ham}, simplifying the problem. We will show in the following section that ground capacitances produce a non-degenerate array mode spectrum whose eigenvectors are eigenstates of parity. The odd parity modes satisfy $\mathcal{N}_{\mu} = 0$, however the even parity modes have $\mathcal{N}_{\mu} \neq 0$. This produces  couplings between the superinductance and parity even array modes. First reported in \cite{di2021efficient}, these new interactions produce important corrections to the circuit Hamiltonian which vary exponentially with mode impedances. 

Since $E^a_J >> E^a_C$, it is common to expand the array mode cosines in \eqref{eq:exact-ham} to second order,
\begin{equation}
    -E^a_J \sum_{m=1}^N \text{cos}\, \theta_m = -E^a_J N + \frac{E^a_J}{2} \theta^T \theta + \mathcal{O}(\theta_m^4)~.
\end{equation}
Performing this expansion and rewriting the $E^b_J$ cosine with trigonometric identities one obtains:
\begin{align}
    \hat H = & 
    - 4 E_C {\partial^2_{\vphi}} - E_J \text{cos}\big( \vphi - \vphi_{\text{ext}} \big)
    + \frac{E_L}{2} \vphi^2 \nonumber \\
    &
    + \sum_{\mu=1}^{N-1} \Big(-4 E_{\mu} {\partial^2_{\mu}}
    + \frac{E^a_J}{2} \Theta^2_{\mu} \Big) \nonumber \\
    & - E_J\text{cos}\big( \vphi - \vphi_{\text{ext}} \big)\Big[ \text{cos}\big(\sum_{\mu}\Theta_{\mu} \mathcal{N}^{1/2}_{\mu} \big) -1 \Big] \nonumber \\
    & + E_J\text{sin}\big( \vphi - \vphi_{\text{ext}} \big)\text{sin}\big(\sum_{\mu}\Theta_{\mu} \mathcal{N}^{1/2}_{\mu} \big),
\end{align}
where, $E_J = E^b_J$, $E_L = E^a_J / \mathcal{N}$. The first line is \eqref{eq:simp-ham}, with the relations between $E_C, E_J, E_L$ and microscopic parameters now obtained. The remaining terms in the Hamiltonian describe array modes and inter-mode couplings~\cite{hazard2019nanowire}.

Examining the Hamiltonian, one can see how the parameters of fluxonium are renormalized through various effects. For example the  inductive energy is given by $E_L = E^a_J/\mathcal{N}$, and only if the superinductance mode has a flat voltage profile does $\mathcal{N} = N$, producing the usual relationship $E_L = E^a_J/N$. In general $\mathcal{N}\leq N$, raising the inductive energy. The quartic (and higher order) couplings that we have dropped further renormalize $E_L$.

The renormalization of other couplings, such as $E_J$, can be compactly expressed in terms of the plasma frequencies and zero-point fluctuations:
\begin{align}
    \hbar \omega_{\mu} & = \Big( 8 E_{\mu} E^a_J \Big)^{1/2} =  \Big( 4 E^a_J e^2 \lambda_{\mu} \Big)^{1/2} \nonumber \\
    \Theta_{\mu, \text{zpf}} & = \Big( 2 E_{\mu}/ E^a_J \Big)^{1/4} = \Big( e^2 \lambda_{\mu} / E^a_J\Big)^{1/4} 
\end{align}
Assuming array modes occupy their ground state, 
\begin{equation}
\label{eq:fluct}
    \langle \text{cos}\big(\sum_{\mu,b}\Theta_{\mu} \mathds{V}_{\mu b} \big) -1 \rangle  \simeq -\frac{1}{2}\sum_{\mu} \mathcal{N}_{\mu} \Theta^2_{\mu, \text{zpf}} \, \nonumber .
\end{equation}
Thus to leading order in fluctuations $E_J$ is renormalized as:
\begin{equation}
\label{eq:ej-renorm}
    E_J \rightarrow E_J \Big(1-\frac{1}{2} \sum_{\mu } \mathcal{N}_{\mu} \Theta^2_{\mu, \text{zpf}} \Big)~.
\end{equation}
The eigensystem of $\mathds{C}^{-1}$ is thus seen to affect many properties of fluxonium: the charging energies of array modes, the renormalization of Hamiltonian couplings, 
as well as dispersive couplings not discussed here \cite{Viola_2015}.  
The eigensystem has been analytically solved in several limits. The case $C^b = 0$ was solved in \cite{masluk2012microwave}, where it was shown that the eigenvectors of $\mathds{C}^{-1}$ are plane waves and analytic expressions for the eigenvalues were found. The case $C^a_g = C^b_g = 0$ was solved in \cite{symmetries-and-collective}, and later work included some effects of charge noise \cite{PhysRevB.102.014512}. Non-zero $C^b$ and $C^a_g, C^b_g$ complicates the theory. Leading order interactions between the superinductance and array modes were computed in \cite{Viola_2015}. Recent work has developed a semi-analytical solution for array modes which holds in the presence of ground capacitances and disorder, and performs tensor network calculations of the resulting quantum many-body system \cite{di2021efficient}.


In this work we derive analytic formulas for the eigensystem when $C^a,C^b,C^a_g,C^b_g \neq 0$. In addition to containing as limiting cases the results of \cite{masluk2012microwave,symmetries-and-collective}, our formulas are useful for understanding physics that emerges from heavily grounded or very long arrays, where the eigensystem of $\mathds{C}^{-1}$ deviates from previous results substantially. 


\section{Eigensystem of $\mathds{C}^{-1}$ }\label{esys}
In this section we derive exact expressions for the eigensystem of $\mathds{C}^{-1}$.
We do so in two steps: we find a simple representation of $\mathds{C}^{-1}$, then  diagonalize it.
Recall that
\begin{equation}
\mathds{C} = C_1 + C_2 + C_3 + C_4.
\end{equation}
It is beneficial to consider a partial sum of $\mathds{C}$, in particular the inverse $(C_2 + C_3 + C_4)^{-1}$. It is straightforward to verify that,
\begin{equation}
\label{eq:cap-to-lap}
    \frac{1}{C_2+C_3+C_4} = \frac{-\Delta(\epsilon,\epsilon')}{C^a_g}~,
\end{equation}
where
\begin{equation}
    -\Delta(\epsilon, \epsilon'):=
    \begin{bmatrix}
        1+\epsilon & -1 & {} & {} & -\epsilon' \\
        -1 & 2 & -1 & {} & {}  \\
        {} & \ddots & \ddots & \ddots & {} \\
        {} & {} & -1 & 2 & -1 \\
        -\epsilon' & {} & {} & -1 &  1 +\epsilon
    \end{bmatrix}~,
\end{equation}
is an $N\times N$ matrix and
\begin{equation}\label{epsepsprime}
    \epsilon = \frac{C^a_g}{C^b_g}\frac{C^b + C^b_g}{2 C^b + C^b_g}, ~~\mathrm{and}~~
    \epsilon' = \frac{C^a_g}{C^b_g}\frac{C^b}{2 C^b + C^b_g}~.
\end{equation}
We observe that $(C_2+C_3+C_4)^{-1}$ represents a discrete Laplace operator.
The $(1,1)$, $(N,N)$, $(1,N)$, and $(N,1)$ entries of the matrix encode boundary conditions satisfied by the corresponding continuous function on the interval. These boundary conditions will in general fix a linear combination of the solution and its derivative at the left/right endpoints (i.e., Robin-Robin boundary conditions). For example, when $\epsilon=1$, and $\epsilon'=1$, the matrix represents a 1D Laplace operator with periodic boundary conditions.  It is conceivable that non-nearest-neighbor couplings will lead to inverse capacitance matrices of a similar form, but with further off-diagonal bands. We speculate that the methods presented in this section, which rely primarily on the structure of banded matrices, may be used to also solve these problems. 

The eigensystem of $\mathds{C}^{-1}$ can be derived by noting that if, for some vector $v$ and scalar $\lambda$,
\begin{equation}
\frac{1}{C_2 + C_3+C_4} v = \lambda v ~,
\end{equation}
then
\begin{equation}
\label{eq:slip}
\frac{1}{C_1 + C_2 + C_3+C_4} v = \frac{\lambda}{1+ C^a \lambda} v ~,
\end{equation}
which holds since $C_1$ is a multiple of the identity operator. Thus, is it enough to compute the spectrum of $(C_2 + C_3 + C_4)^{-1}$, and hence $-\Delta(\epsilon,\epsilon')$,  to compute that of $\mathds{C}^{-1}$. It is important to note that while $1/(C_2 + C_3 + C_4)$ is sparse, $1/(C_1 + C_2 + C_3 + C_4)$ is dense, as can be seen by Taylor expanding in $C_1$ or numerically computing examples at small $N$. Thus this seemingly trivial step is actually quite important for the analysis. 

Before proceeding we would like to point out that Eq. \ref{eq:cap-to-lap} itself is useful. The capacitance matrix itself is dense, and numerically computing its eigensystem becomes costly for long arrays: Eq. \ref{eq:cap-to-lap} shows the eigensystem can be computed from a sparse matrix.

We will now compute the eigensystem of $-\Delta(\epsilon,\epsilon')$. We will distinguish the eigenvalues of $-\Delta(\epsilon,\epsilon')$ and $\mathds{C}^{-1}$, writing the former as $\ell$ and the latter as $\lambda$. For every eigenvalue $\ell$ of $-\Delta(\epsilon,\epsilon')$, the corresponding eigenvalue of $\mathds{C}^{-1}$ is $\lambda = (C^a + C^a_g / \ell)^{-1}$. The eigenvectors of the two matrices are equal and we denote these as $v$.

The eigensystem of $-\Delta(\epsilon,\epsilon')$ can be derived by exploiting reflection symmetry of the fluxonium circuit about its midpoint, which is present since we are considering a differential fluxonium circuit without array disorder. This symmetry is manifested in the fact that $-\Delta(\epsilon,\epsilon')$
commutes with 
\begin{equation}
    \mathds{F} := 
    \begin{bmatrix}
        {}  & {} &  1 \\
        {}  & \iddots  & {} \\
        1  & {} & {}  \\
    \end{bmatrix}~,
\end{equation}
which implements the reflection.
Since $\mathds{F}^2 = \mathds{1}$, the following matrices are projection operators
\begin{equation}
    P_{\pm} = \frac{1}{2} \big( \mathds{1} \pm \mathds{F} \big) ~,
\end{equation}
which allows $-\Delta(\epsilon,\epsilon')$ to be decomposed over subspaces as follows: 
\begin{equation}
\label{eq:proj}
    -\Delta(\epsilon,\epsilon') = \sum_{i = \pm} P_i \Big\{ -\Delta(\epsilon,\epsilon') \Big\} P_i~.
\end{equation}
We call the $P_+$ subspace ``parity even" and the $P_-$ subspace ``parity odd" (the voltage profiles of $P_+$ \& $P_-$ modes are respectively odd \& even).
Each term on the right-hand-side of \eqref{eq:proj} can be simplified:
\begin{equation}\label{epsilonplusminus}
    P_{\pm} \Big\{ -\Delta(\epsilon,\epsilon') \Big\} P_{\pm} = -\Delta(\epsilon_{\pm},0) P_{\pm} ~,
\end{equation}
where
\begin{align}
    \epsilon_+ & := \epsilon - \epsilon' = \frac{C^a_g}{2C^b + C^b_g} \nonumber \\
    \epsilon_- & := \epsilon + \epsilon' = \frac{C^a_g}{C^b_g}~.
\end{align} 
These expressions show that even modes depend on $C^b$ while odd ones do not, which is expected because only even modes drop in voltage across the small junction.

Denoting the eigenvectors of $-\Delta(\epsilon,\epsilon')$ which are members of the $P_+$, $P_-$  subspaces as $v_+(\ell_{+})$, $v_-(\ell_-)$ respectively, \eqref{epsilonplusminus} then implies
\begin{align}
    -\Delta(\epsilon,\epsilon') v_{\pm}(\ell_{\pm}) & = -\Delta(\epsilon_{\pm},0)v_{\pm}(\ell_{\pm}) \nonumber \\
    & =\ell_{\pm} v_{\pm}(\ell_{\pm})~.
\end{align}
Thus it suffices to attain the eigensystems of the matrices $-\Delta(\epsilon_\pm,0)$. The characteristic polynomials of $-\Delta(\epsilon_\pm,0)$ are in fact doubly-convex combinations in $\epsilon_\pm$ (see Appendix \ref{derive-char-poly} for a derivation). 
The eigenvalues of $-\Delta(\epsilon_{\pm},0)$
are respectively the roots of these two degree $N$ polynomials:
\begin{align}
\label{eq:char-poly}
    \mathcal{P}_{\pm} & = (1-\epsilon_{\pm})^2 \Big[(2\alpha_{\pm} - 2) \mathrm{U}_{N-1}(\alpha_{\pm})\Big]  + \epsilon_{\pm}^2 \Big[ \mathrm{U}_N(\alpha_{\pm})  \Big] \nonumber \\
   & + 2(1-\epsilon_{\pm})\epsilon_{\pm} \Big[\beta_{\pm}^{-1}\mathrm{T}_{2N+1}(\beta_{\pm})\Big]~, \\
   &\mathrm{for}\quad \alpha_{\pm} := 1 - \ell_{\pm}/2, ~~ \beta_{\pm} := (1-\ell_{\pm}/4)^{1/2} ~,\nonumber
\end{align}
where $\mathrm{T}_k$ \& $\mathrm{U}_k$ are $k^{\text{th}}$ order Chebyshev polynomials of the first \& second kind \cite{DLMF_NIST,chebyshev-det,chebyshev-det2} (the $\mathrm{T}_{2N+1}$ term is a polynomial in $\lambda$, even though the argument involves a square root, see Appendix \ref{derive-char-poly}). The $m^{\text{th}}$ component of the (non-normalized) eigenvector corresponding to eigenvalue $\ell_{\pm}$ can be expressed as 
\begin{align}
\label{eq:evecs}
    [v_{\pm}(\ell_{\pm})]_m & =  \epsilon_{\pm} \Big[\mathrm{U}_{m-1}(\alpha_{\pm})\Big] \nonumber \\
    & + (1-\epsilon_{\pm})\Big[\beta_{\pm}^{-1}\mathrm{T}_{2m-1}(\beta_{\pm})\Big]~,
\end{align}
where $m=1,\dots,N$. \eqref{eq:char-poly}, \eqref{eq:evecs} give a complete description of eigenvalues/vectors of the inverse capacitance matrix (by making use of \eqref{eq:slip} \& \eqref{eq:cap-to-lap}).


Before continuing we address an important detail. Each of the polynomials $\mathcal{P}_+$ and $\mathcal{P}_-$ has $N$ roots: together they have $2N$, which is double the number of eigenvalues. To obtain the $N$ eigenvalues of $-\Delta(\epsilon,\epsilon')$, one computes all $2N$ roots, plugs them into the eigenvector formula \eqref{eq:evecs}, and keeps only eigenvalues which produce eigenvectors with the correct symmetry. In particular, the $N$ roots of $\mathcal{P}_+$ are to be plugged into the $v_+(\ell_+)$ formula, and only those eigenvalues which produce even vectors are kept. Similarly the $N$ roots of $\mathcal{P}_-$ are plugged into $v_-(\ell_-)$, and only those eigenvalues which produce odd vectors are kept. This procedure results in $N$ eigenvalues/vectors, and these constitute the eigensystem of $-\Delta(\epsilon,\epsilon')$. 

With the correct roots obtained, there is an alternative, simpler formula for the eigenvectors. Defining a vector-valued function $\tilde{v}(x)$ with components
\begin{equation}\label{eq:mystery}
    \big[ \tilde{v}(x) \big]_m=  \text{cos}\Big((2m-1 )\text{arcsin}\sqrt{x/4} \Big),
\end{equation}
for $m=1,\dots,N$, then a non-normalized eigenvector corresponding to even/odd eigenvalue $\ell_{\pm}$ is
\begin{equation}\label{eq:mystery2}
    P_{\pm} \tilde{v}(\ell_{\pm})~.
\end{equation}
The eigenvectors are thus seen to be even and odd superpositions of plane waves. With these expressions one can show that for even modes
\begin{equation}\label{eq:n-mus}
    \mathcal{N}_{\mu} 
    = 
    \frac{8\, \text{sin}\Big(N \text{csc}^{-1}\sqrt{\frac{4}{\ell_+} } \Big)^2 / \ell_+ }{N + \text{csc}\Big(2 \text{csc}^{-1}\sqrt{\frac{4}{\ell_+} }\Big) \text{sin}\Big(2 N \text{csc}^{-1} \sqrt{\frac{4}{\ell_+} } \Big)} ~,
\end{equation}
for the $\mu^{\text{th}}$ even eigenvalue $\ell_+$ (including the zeroth eigenvalue), and $\mathcal{N}_{\mu} = 0$ for all odd modes. It is useful to expand $\mathcal{N}_{\mu}$ when ground capacitances are small. In this case $\epsilon_+ \approx 0$, and the characteristic polynomial $\mathcal{P}_{+} \approx (2\alpha_{+} - 2) \mathrm{U}_{N-1}(\alpha_{+})$. The roots of this polynomial are $0, 4\, \text{sin}^2(\mu \pi / 2N)$, for $\mu = 1, \dots, N-1$, whose positions will be perturbed by ground capacitances to $0+\delta_0, 4\, \text{sin}^2(\mu \pi / 2N)+ \delta_{\mu}$. One finds
\begin{align}\label{eq:n-mus-approx}
    \mathcal{N} & = N + \frac{\delta_0^2}{720}\big(-4N + 5N^3 - N^5\big) + \mathcal{O}(\delta_0^3) \nonumber \\
    \mathcal{N}_{\mu} & = \delta_{\mu}^2 \frac{N}{32}\text{csc}\Big(\frac{\mu \pi}{2N}\Big)^{4}\text{sec}\Big(\frac{\mu \pi}{2N}\Big)^{2} + \mathcal{O}(\delta_{\mu}^3)~,
\end{align}
for even modes, while for odd modes $\mathcal{N}_{\mu}$ always vanishes.

\begin{figure}[t]
    \centering
    \includegraphics[width=0.45\textwidth]{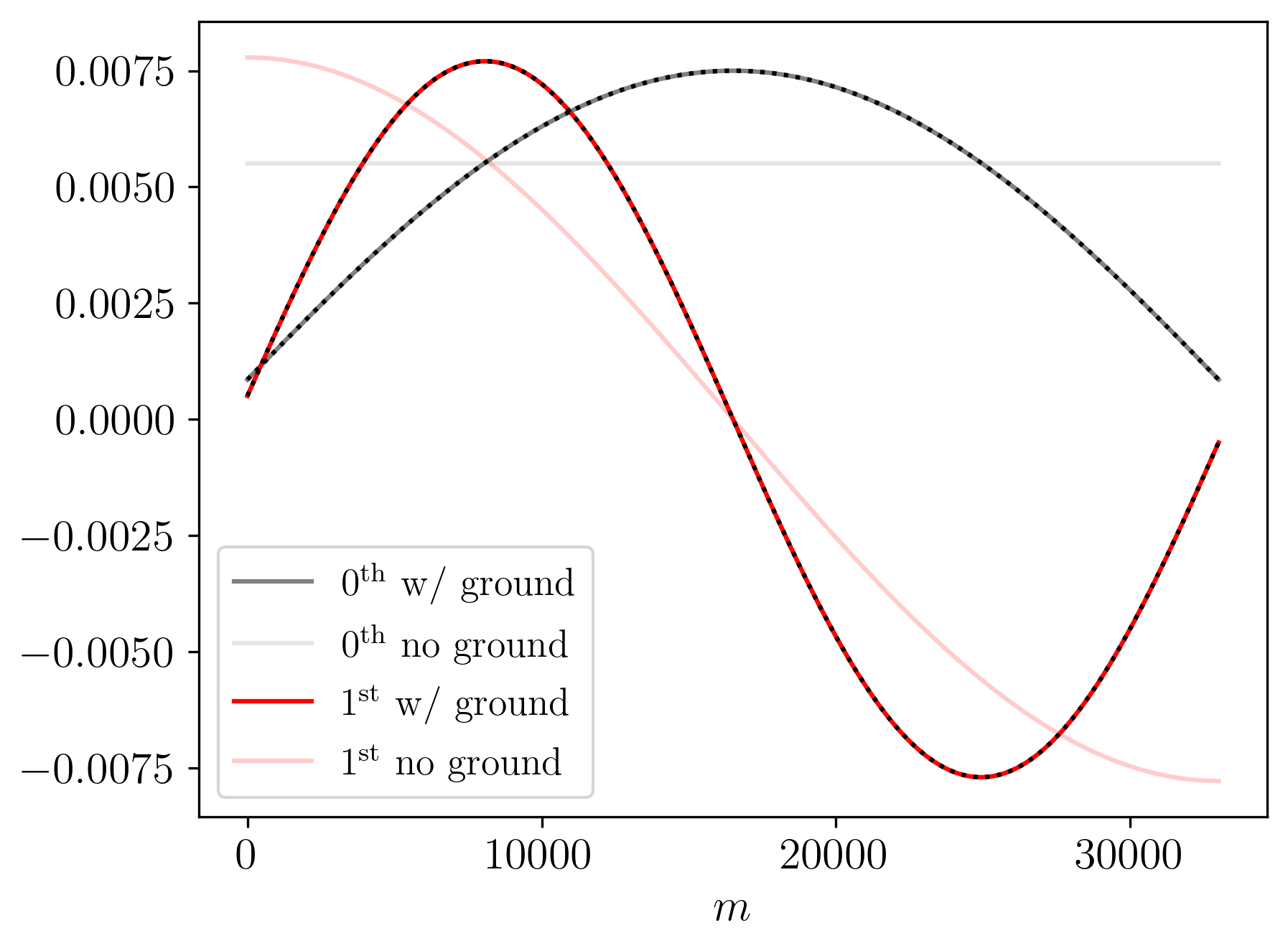}
    \caption{Two lowest eigenvectors of an $N=33,000$ array with charging energies of Eq. \ref{eq:params}. Opaque/transparent curves include/neglect ground capacitance respectively. Dotted black lines result from numerically diagonalizing the inverse capacitance matrix.}
    \label{fig:vectors}
\end{figure}

We now provide a numerical demonstration of our formulas in Fig. \ref{fig:vectors}. Motivated by an experiment reaching $N = 33,000$ \cite{Kuzmin_2019}, we consider a very long $N=33,000$ array, where ground effects are large, with capacitances 
\begin{equation}
\label{eq:params}
    (C^a,C^b,C_g^a,C_g^b) = (19.37,5.23,0.01,3.87) ~[\text{fF}] \, ,
\end{equation}
taken from Appendix F of \cite{symmetries-and-collective} (we convert the reported charging energies to capacitances via $E_C = e^2/2C$).
We solve for the lowest two modes of $\mathds{C}^{-1}$ using our exact result and compare against exact diagonalization. The dotted black lines in the figure are obtained by numerically diagonalizing  the inverse capacitance matrix, and their agreement with the red \& gray curves, obtained from \eqref{eq:evecs}, demonstrates our formulas are correct. The transparent curves are the approximate eigenvectors of \cite{symmetries-and-collective}, obtained by dropping ground capacitances (i.e. setting $C_3 = C_4 = 0$). The difference between transparent and opaque curves indicates that ground capacitances substantially modify mode profiles in long arrays.

We conclude this section by showing how the preceding results apply to grounded (rather than differential) fluxonium devices. Consider a grounded fluxonium device with $N$ array junctions, Lagrangain given by Eqs.
\ref{eq:lagrangian}-\ref{eq:kinetic}, and capacitances $C^a, C^b, C^a_g, C^b_g$. Then all that changes in the derivation of the quantum Hamiltonian of this circuit is that $\dot \tau = 0$ from the outset. The effect of this change is to eliminate $C_4$ from the capacitance matrix, so that $\mathds{C} = C_1 + C_2 + C_3$. The eigensystem of this new capacitance matrix can be solved with similar methods as the differential case. One finds that
\begin{equation}
\label{eq:grounded-cap-mat}
    \frac{1}{C_2 + C_3} = \frac{1}{C^a_g}
    \begin{bmatrix}
        1 & -1 & {} & {} & {} \\
        -1 & 2 & -1 & {} & {}  \\
        {} & \ddots & \ddots & \ddots & {} \\
        {} & {} & -1 & 2 & -1 \\
        {} & {} & {} & -1 &  1 + \delta
    \end{bmatrix}  := \frac{-\Delta(\delta)}{C^a_g} ~,
\end{equation}
where $\delta = C^a_g / (C^b + C^b_g)$ and where we have defined a new $N \times N$ matrix $-\Delta(\delta)$. 

The eigensystem of \eqref{eq:grounded-cap-mat} can be obtained by noticing that it is embedded in that of a differential device with $2N$ array junctions and small junction capacitance $C^b/2$. In this case,
\begin{equation}
    -\Delta(\epsilon_+,0) P_+ = 
    \frac{1}{2}\begin{bmatrix}
        -\mathds{F} \Delta(\delta) \mathds{F} & -\mathds{F} \Delta(\delta) \\
        - \Delta(\delta) \mathds{F}  & -\Delta(\delta)
    \end{bmatrix}~,
\end{equation}
where it is understood that the matrix on the left hand side is $2N\times 2N$ and the blocks on the left are $N\times N$. The even eigenvectors of the $2N$-sized system can be written as $v_{2N} = \begin{bmatrix} \mathds{F} v_N & v_N \end{bmatrix}$, and one can show that $-\Delta(\epsilon_+,0) P_+ v_{2N} = \lambda v_{2N}$ implies $-\Delta(0,\delta)v_N = \lambda v_N$. Thus the eigenvectors of the grounded system with $N$ array junctions and small junction capacitance $C^b$ are the ``first half" of the eigenvectors of the differential system with $2N$ array junctions and small junction capacitance $C^b/2$, and the eigenvalues between the two cases are the same. 

This same conclusion can be reached from symmetry arguments \cite{PhysRevX.13.031035,PhysRevX.9.041041}. The even modes of a differential device with $2N$ array junctions have zero voltage at the mid point of the circuit. Rewriting the small junction capacitor as two equal capacitors in series, and noting the reflection symmetry about the midpoint shows that, from the point of view of the even modes, the first $N$ junctions of the circuit is indistinguishable from a grounded device.

\section{Approximation Scheme}\label{approx}
\begin{figure*}[t]
\begin{center}
	\subfloat[\label{fig:comparison}]{
            \includegraphics[width=0.4\textwidth]{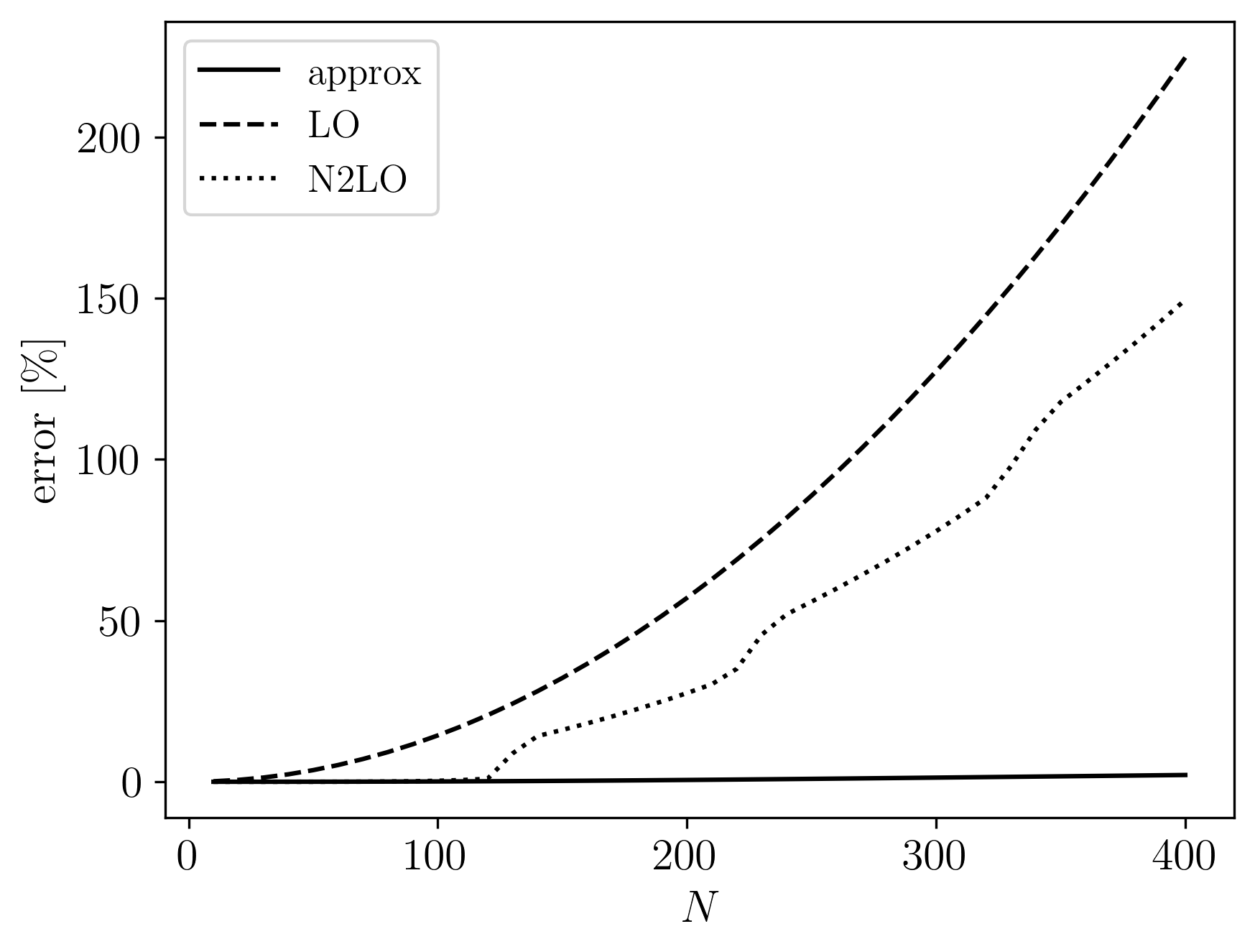}
            }
	\subfloat[\label{fig:sweep-across-N}]{
            \includegraphics[width=0.4\textwidth]{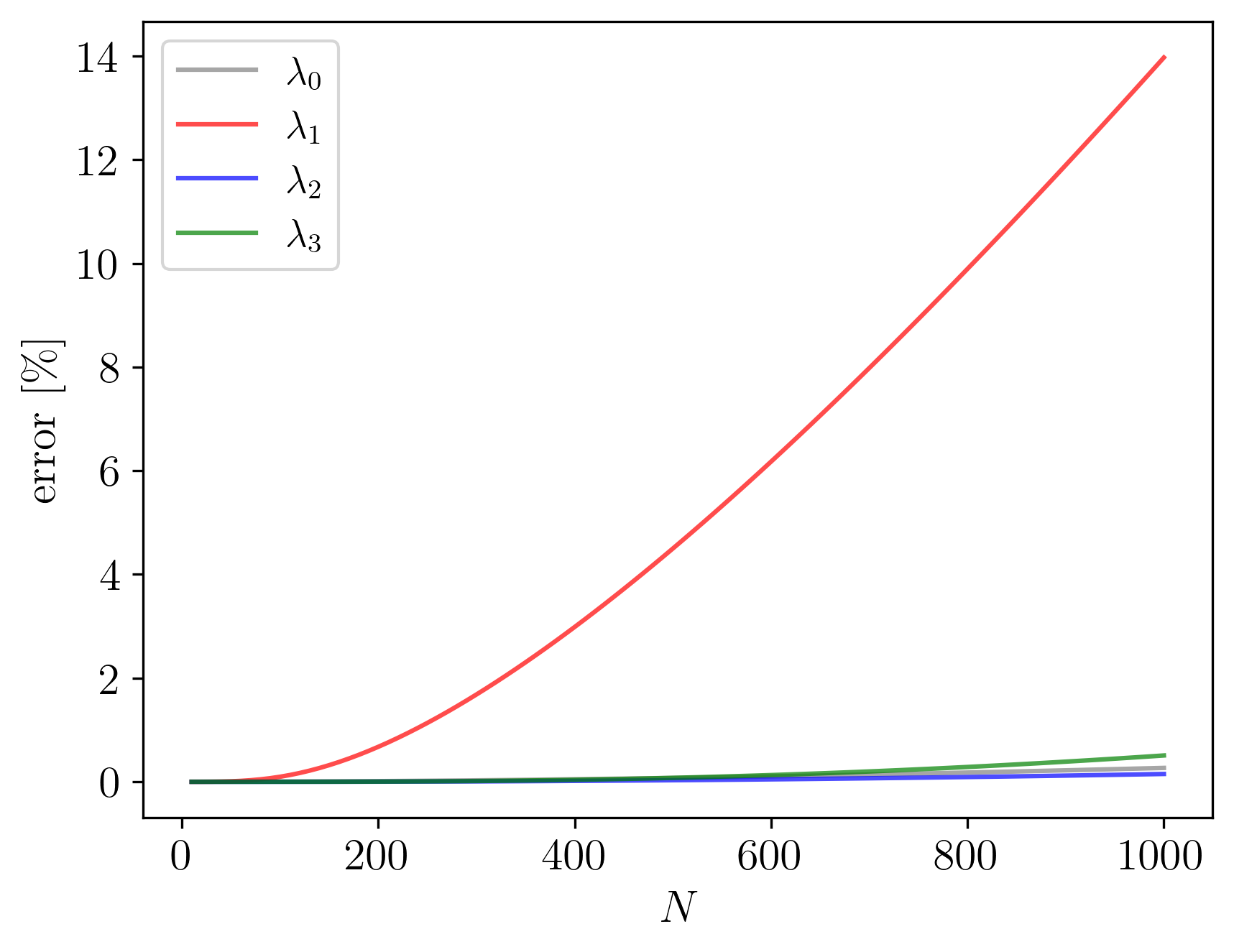}
            }\\
        \subfloat[\label{fig:approx-vectors}]{
            \includegraphics[width=0.4\textwidth]{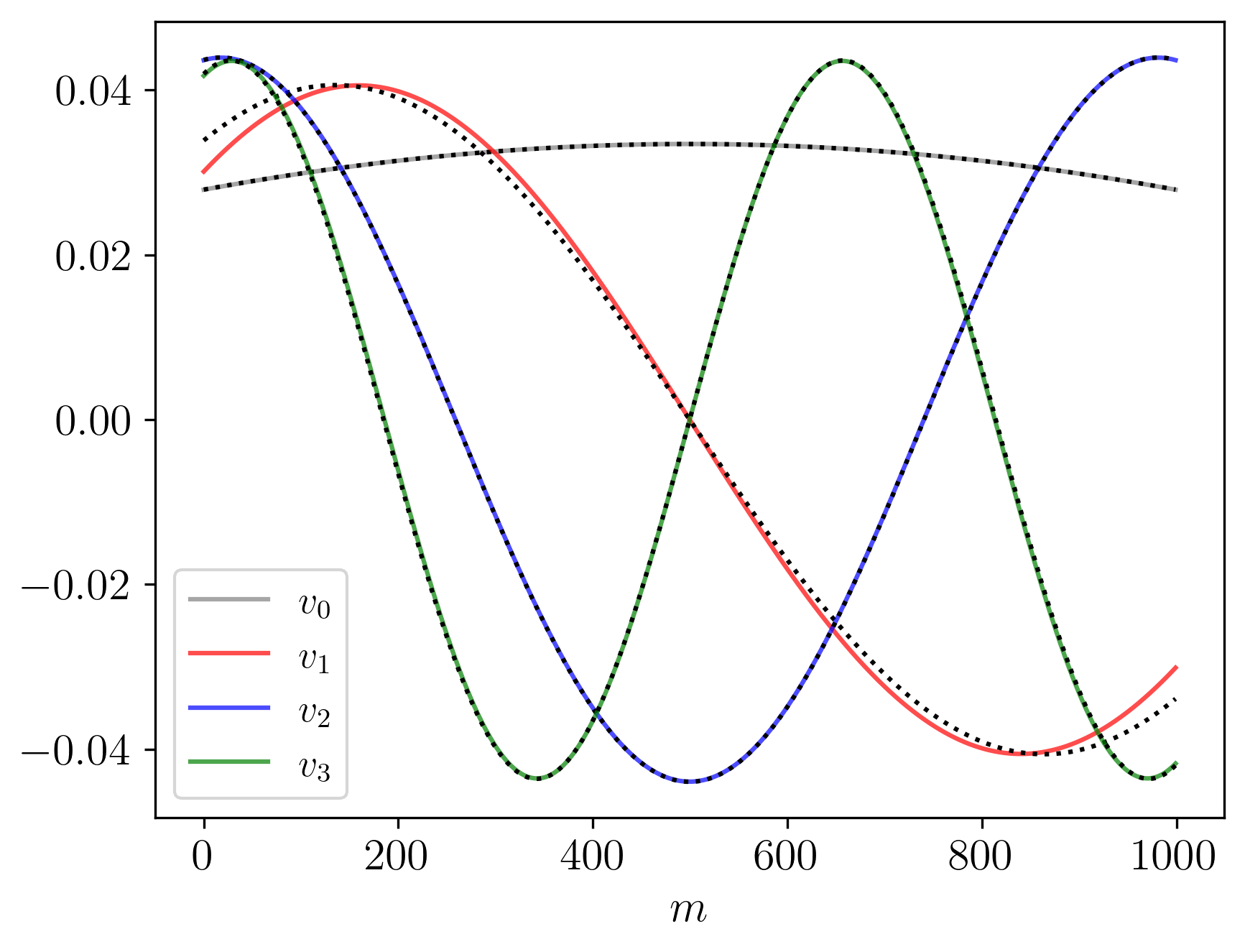}
            }
	\subfloat[\label{fig:sweep}]{
            \includegraphics[width=0.4\textwidth]{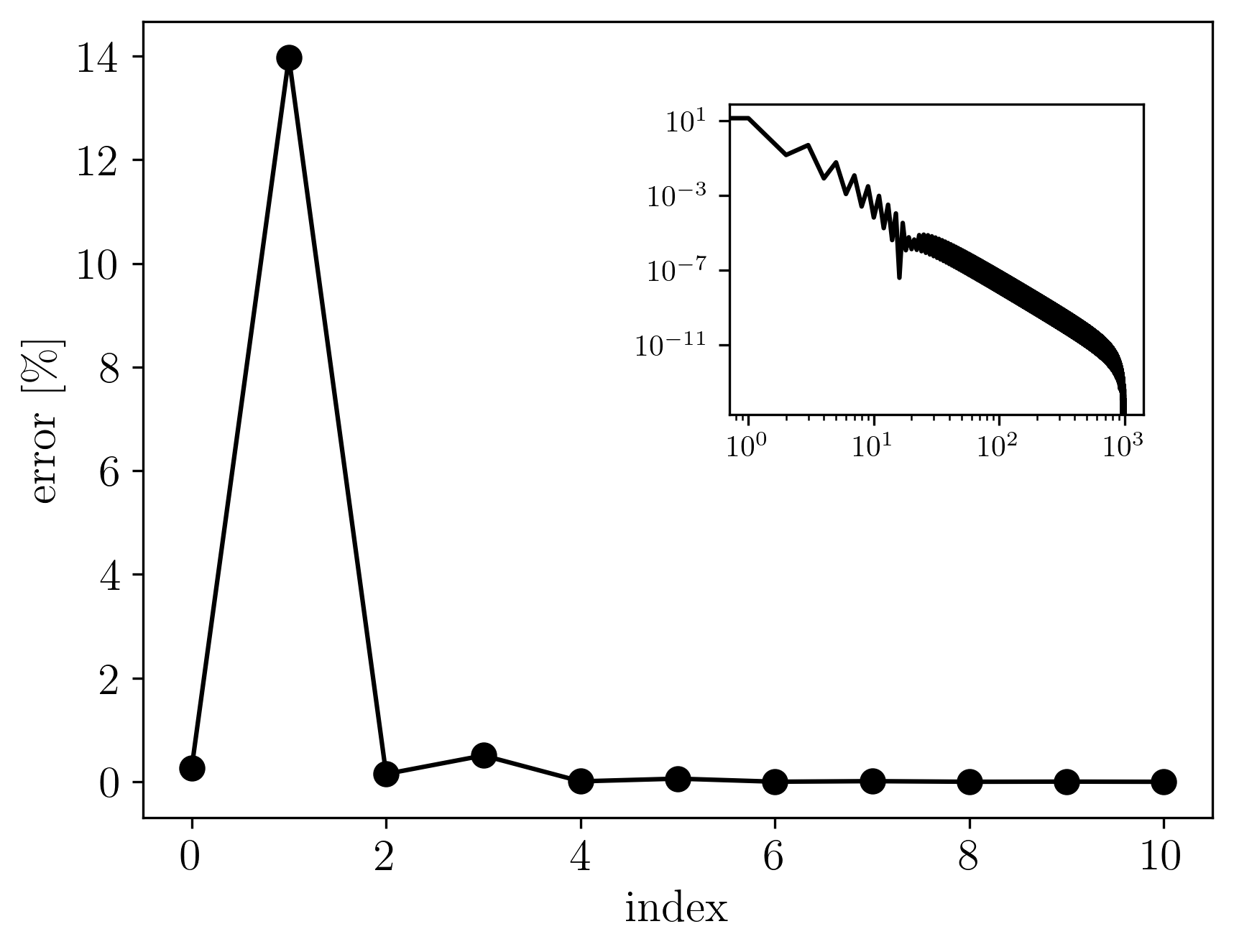}
            }
\end{center}
    \caption{Approximation scheme performance. 
    \protect\subref{fig:comparison} 
    Fractional error of $\lambda_2$ in different approximation schemes. Our scheme is labelled ``approx", while LO and N2LO indicate leading- \& next-to-next-to-leading-order perturbation theory.
    \protect\subref{fig:sweep-across-N}
    Fractional error of our approximation scheme for the lowest four eigenvalues of $\mathds{C}^{-1}$ as a function of the number of array junctions $N$.
    \protect\subref{fig:approx-vectors}
    Exact and approximate eigenvectors of four lowest modes at $N=1,000$. Colored lines are the approximate eigenvectors while the dotted black lines are the exact eigenvectors.
    \protect\subref{fig:sweep}
    Fractional error of eigenvalues at $N=1,000$ in our approximation scheme. The horizontal axis indexes eigenvalues. Points show the lowest eigenvalues while the inset shows the whole spectrum.
    }
    \label{fig:test}
\end{figure*}

The formulas of \eqref{eq:char-poly}, \eqref{eq:evecs} for the eigensystem of $\mathds{C}^{-1}$ are exact. However it may be useful to work with approximate eigensystems with simple expressions requiring no root finding. We presently provide such a scheme. The approximate eigenvalues of $\mathds{C}^{-1}$ are
\begin{align}
\label{eq:eval-est}
    \lambda_0 & =  \big[
    C^a + C^a_g /\ell_0
    \big]^{-1} \nonumber \\
    \lambda^{\text{even}}_{\mu} & = \big[ C^a + C^a_g /\ell^{\text{even}}_{\mu} 
    \big]^{-1} \nonumber \\
    \lambda^{\text{odd}}_{\mu} & = \big[ C^a + C^a_g / \ell^{\text{odd}}_{\mu} 
    \big]^{-1} ~,
\end{align}
where
\begin{align}
\label{eq:deltas}
    \ell_0 & =  \big[ \big(\frac{N^2}{12} - \frac{N}{4} + \frac{1}{6}\big)+ N C^b/C^a_g + \frac{N}{2}C^b_g/C^a_g \big]^{-1}  \nonumber \\
    \ell^{\text{even}}_{\mu} & = 4\,  \text{sin}^{2}\big(\frac{\mu \pi}{2N}\big) + \frac{4}{N}\text{cos}^{2}\big(\frac{\mu \pi}{2N}\big)\frac{C^a_g }{2C^b + C^b_g}\nonumber \\
    \ell^{\text{odd}}_{\mu} & = 4\,  \text{sin}^{2}\big(\frac{\mu \pi}{2N}\big) + \frac{4}{N} \text{cos}^{2}\big(\frac{\mu \pi}{2N}\big)\frac{C^a_g}{C^b_g}~,
\end{align}
and the approximate eigenvectors of $\mathds{C}^{-1}$ are
\begin{align}
    v_0 & = P_{+} \tilde{v}(\ell_0) \nonumber \\
    v^{\text{even}}_{\mu} & = P_{+} \tilde{v}(\ell^{\text{even}}_{\mu}) \nonumber \\
    v^{\text{odd}}_{\mu} & = P_{-} \tilde{v}(\ell^{\text{even}}_{\mu})~.
\end{align}
where $\mu=1,\dots,N-1$. The $\mathcal{N}, \mathcal{N}_{\mu}$ corresponding to these approximate eigenvectors can be obtained by substituting $\ell_0,\ell^{\text{even}}_{\mu}$ into \eqref{eq:n-mus} or approximated by \eqref{eq:n-mus-approx}. For the odd modes $\mathcal{N}_{\mu} = 0$.

We arrive at this approximation scheme in the following way. First, the approximate eigenvectors are simply the exact eigenvectors of $-\Delta(\epsilon,\epsilon')$ (and therefore $\mathds{C}^{-1}$) evaluated at approximate eigenvalues. This choice retains the correct symmetry and wave-like profile of the exact solution, but uses an approximate wave-number. To estimate the eigenvalues of $\mathds{C}^{-1}$, we leverage the eigenvectors of \cite{symmetries-and-collective}, namely:
\begin{align}
\label{eq:zeroth-order-evals}
    \big( \mathbf{v}_0 \big)_m & = N^{-1/2} \nonumber \\
    \big( \mathbf{v}_{\mu} \big)_m & =\sqrt{\frac{2}{N}} \text{cos}\big(\frac{\pi \mu}{N}(m-1/2)\big)~,
\end{align}
which can be shown to be the limit of the exact eigenvectors  as ground capacitances go to zero. We then separate two cases: the lowest eigenvalue and the rest of them. For the latter we compute
\begin{equation}
\label{eq:sandwich}
    \mathbf{v}_{\mu}^{\dagger} \Big[ \frac{-\Delta(\epsilon,\epsilon,\epsilon',N)}{C^a_g}\Big] \mathbf{v}_{\mu}~,
\end{equation}
which is an estimate of the $\mu^{\text{th}}$ eigenvalue of $(C_2 + C_3 + C_4)^{-1}$. We then use \eqref{eq:slip} to relate this eigenvalue estimate to one for the full $\mathds{C}^{-1}$. Depending whether $\mathbf{v}_{\mu}$ is an even or odd mode, either $\epsilon_+ = C^a_g/(2C^b + C^b_g)$ or $\epsilon_- = C^a_g/C^b_g$ appears in the calculation, and this produces the different dependences in \eqref{eq:deltas}. The dependence/independence of even/odd vectors on $C^b$ is expected from symmetry, and it is a strength of our scheme this feature is captured. Futhermore, since ground capacitances are typically much smaller than junction capacitances, $\epsilon_+ << \epsilon_-$, and one therefore expects better estimates for the even modes than the odd. 

Repeating the above procedure with $\mathbf{v}_0$, one obtains $(C^a+N C^b + C^b_g N/2)^{-1}$ as an estimate of the lowest eigenvalue of $\mathds{C}^{-1}$. A tighter bound can be obtained, however, by observing that the lowest eigenvalue of $\mathds{C}^{-1}$ is equal to the inverse of the highest eigenvalue of $\mathds{C}$. Thus the lowest eigenvalue of $\mathds{C}^{-1}$ is bounded from above by $\lambda_0 := \big( v_0^{\dagger} \mathds{C} v_0 \big)^{-1}$, which is equal to the expression in \eqref{eq:eval-est}. Since $\lambda_0 < (C^a+N C^b + C^b_g N/2)^{-1}$, it is a tighter bound and we use it. Previous work arrived at this same estimate \cite{symmetries-and-collective}. Finally, we have shown that solving for the roots of the exact solution \eqref{eq:char-poly} to first order in $\epsilon_{\pm}$ yields the same approximate eigenvalues as the present method.

We now give an example to illustrate the utility of the approximation scheme. For the rest of this section we consider systems with variable $N$, but with capacitances fixed to those listed in \eqref{eq:params}, which we reproduce here
\begin{equation}
    (C^a,C^b,C_g^a,C_g^b) = (19.37,5.23,0.01,3.87) ~[\text{fF}] \, .
\end{equation}
In Fig. \ref{fig:comparison} we plot the fractional error of the first even parity array mode $\lambda_2$ for $N\leq 400$, where fractional error is defined as $100 \times |  (\lambda_{\text{exact}}-\lambda_{\text{approx}})/ \lambda_{\text{exact}}  |$. The estimate is quite good, with less than $1\%$ error by $N=400$. In the figure we also compare to perturbation theory, which approximates the full inverse capacitance matrix by the first few terms in the geometric series
\begin{equation*}
    \frac{1}{C_1+C_2+C_g} = \sum_{k=0}^{\infty} (-1)^{k} \frac{1}{C_1+C_2}  \Big[ C_g \frac{1}{C_1+C2}\Big]^k~,
\end{equation*}
where $C_g = C_3 + C_4$ \cite{symmetries-and-collective, Viola_2015}. In the figure ``LO" and ``N2LO" means to cut off the infinite sum at $k=0,2$, respectively. At leading order the superinductance mode eigenvalue equals $1/(C^a + N C^b)$ while the array modes eigenvalues are degenerate with eigenvalue $1/C^a$. We skip NLO $(k=1)$ because the series exhibits an alternating behavior which renders the NLO estimate quite poor. Fig. \ref{fig:comparison} shows our approximation scheme is more accurate than perturbation theory. Since our method requires no matrix multiplication it is also cheaper.

In Fig. \ref{fig:sweep-across-N}, we compare $\lambda_0, \lambda_1, \lambda_2, \lambda_3$ to the exact result for $N \leq 1,000$. The estimates are seen to have at worst a $\sim 14\%$ error; this occurs for the first odd parity mode which has an anomalously large error. All other modes are extremely well-estimated, with less than $2\%$ error by $N=1,000$. Fig. \ref{fig:approx-vectors} compares first four  exact and approximate eigenvectors at $N=1,000$. The parity and wave-like nature of eigenvectors are baked into the approximation scheme, with only the estimated wavenumber slightly incorrect. In Fig. \ref{fig:sweep} we fix $N=1,000$ and plot the spectrum of $\mathds{C}^{-1}$. The low modes have the highest error, however as indicated by the inset the entire spectrum is  well-approximated.

We conclude this section by noting that, while this approximation scheme is quite accurate, it does not capture the correct asymptotic behavior of eigenvalues at large $N$ and will become inaccurate at large enough arrays. Computing eigenvalues for array lengths up to $N = 250,000$, we find strong evidence that the asymptotic scaling of eigenvalues is $N^{-2}$, a feature noted previously \cite{Rastelli_2013}. While our superinductance mode estimate scales as $N^{-2}$, the array mode estimates scale as $N^{-1}$. On very large arrays we suggest the following alternative estimate, which has the correct asymptotic scaling:
\begin{align}
    \Big( \mathbf{v}_{\mu}^{\dagger} & \mathds{C} \mathbf{v}_{\mu} \Big)^{-1}  = \big[ C^a + \frac{C^a_g}{4} \text{sin}^{-2}\big(\frac{\pi \mu}{2 N}\big) \nonumber  \\
    & -\frac{ {(1-(-1)^{\mu})^2  C^a_g}^2 }{2 C^b_g + (N-1)C^a_g}\frac{\text{cot}^2\big(\frac{\pi \mu}{2N}\big)\text{csc}^2\big(\frac{\pi \mu}{2N}\big)}{8N}
    \big]^{-1} ~.
\end{align}

\section{Conclusion}\label{conclusion}
In this work we presented an exact solution for the array modes of fluxonium in the absence of array disorder. Array mode energies and spatial profiles are determined by the eigenvalues and eigenvectors of the inverse capacitance matrix, which we have explicitly solved for. The eigenvalues are the roots of a doubly-convex combination of Chebyshev polynomials, while the eigenvectors are plane waves. These results extend known formulae for the array mode spectrum in the absence of ground capacitances \cite{symmetries-and-collective}.  

In the course of developing this solution, it was shown that the inverse capacitance matrix is related to a discrete Laplace operator with Robin-Robin boundary conditions. This intermediate result has practical utility because it shows that while the capacitance matrix itself is dense, its spectrum can be computed from a sparse matrix. 

Reflection symmetry of the circuit about 
its midpoint is essential in our analysis: it  organizes array modes into even and odd parity subspaces and simplifies the mathematical steps required to find the eigensystem. The even array modes were seen to couple to the superinductance mode through the small junction with a magnitude controlled by the array mode normalizations $\mathcal{N}_{\mu}$. 
These couplings produce corrections to the circuit Hamiltonian which vary exponentially with the mode impedances \cite{di2021efficient}, and to assist in their quantification we have provided both exact and approximate expressions for the normalizations. We then related our results for a differential device to a grounded one, showing that the array modes of a grounded device of length $N$ with small junction capacitance $C^b$ are equal to the ``first half" of the even modes of a length $2N$ floating fluxonium with small junction capacitance $C^b/2$. 

We provided a simple approximation scheme for the eigensystem of the inverse capacitance matrix requiring no Chebyshev polynomials. Eigenvalues are approximated by simple analytic expressions while eigenvectors are plane waves. 
For arrays less than a thousand junctions long, all eigenvalues except the first odd array mode are estimated to better than $2\%$. The eigenvector approximation scheme involves substituting into the exact formula an approximate wavenumber, which approximate eigenvectors well. Our scheme outperforms perturbation theory in both speed and accuracy.

The work presented here focuses on only a small aspect of the physics of fluxonium. It will be interesting to explore the consequences of our formulas on large arrays, where rich new physics in the form of interesting phase transitions and useful devices may potentially reside.

\section{Acknowledgements}
SS is supported by the National Research Council. NCW is supported by: the U.S. Department of Energy, Office of Science under grant Contract Numbers DE-SC0011090 and DE-SC0021006, the Simons Foundation grant 994314 (Simons Collaboration on Confinement and QCD Strings), and the U.S. Department of Energy, Office of Science, National Quantum Information Science Research Centers, Co-Design Center for Quantum Advantage under Contract No. DE-SC0012704. . M. H. was supported by an appointment to the Intelligence Community Postdoctoral Research Fellowship Program at the Massachusetts Institute of
Technology administered by Oak Ridge Institute for Science and Education (ORISE) through an interagency
agreement between the U.S. Department of Energy and
the Office of the Director of National Intelligence (ODNI).
This research was funded in part by the U.S. Army Research Office under Award No. W911NF-23-1-0045.

\bibliography{bib}

\appendix
\section{Characteristic polynomial derivation }\label{derive-char-poly}
In this appendix we derive the characteristic polynomial of the $N\times N$ matrix
\begin{equation}
\label{eq:discrete-laplace-operator}
    -\Delta(\epsilon,0) = 
    \begin{bmatrix}
        1+\epsilon & -1 & {} & {} & {} \\
        -1 & 2 & -1 & {} & {}  \\
        {} & \ddots & \ddots & \ddots & {} \\
        {} & {} & -1 & 2 & -1 \\
        {} & {} & {} & -1 &  1 +\epsilon
    \end{bmatrix}~,
\end{equation}
for general $\epsilon, N$. We do this by considering an auxiliary matrix   
\begin{equation}
    -\Delta(\epsilon) :=     
    \begin{bmatrix}
        1 & -1  & {}  & {} & {} \\
        -1 & 2  & -1  & {} & {} \\
        {} &  \ddots & \ddots & \ddots & {} & {} \\
        {} &  {} & -1  & 2 & -1 \\
        {} &  {} & {}  & -1 &  1 +\epsilon
    \end{bmatrix}~,
\end{equation}
explaining at the end how to extend the result to (\ref{eq:discrete-laplace-operator}). Denote this characteristic polynomial by
\begin{equation}
    P(\lambda, \epsilon) := \text{det}\big( -\Delta(\epsilon) - \lambda \mathds{1} \big).
\end{equation}
The recurrence relation for the determinant of a tridiagonal matrix \cite{dets-of-tridiag, Horn_Johnson_1985} gives
\begin{equation}
    P(\lambda,\epsilon) = (1+\epsilon -\lambda) p_{N-1} - p_{N-2},
\end{equation}
where $p_i$ is the determinant of an $i \times i$ matrix:
\begin{equation}
    p_i = \text{det}
    \begin{bmatrix}
        1-\lambda & -1 {} & {} & {} &{}\\
        -1 & 2-\lambda & -1 &{} & {} & {}\\
        {} & {\ddots} & \ddots & {\ddots} & {}\\
        {} & {} &  -1 & 2-\lambda & -1 \\
        {} & {} & {}  & -1 & 2-\lambda 
    \end{bmatrix}.
\end{equation}
Each $p_i$ is a difference of Chebyshev poylnomials of the second kind \cite{chebyshev-det}, 
\begin{equation}
p_i = \mathrm{U}_{i-1}(\alpha) - \mathrm{U}_{i-2}(\alpha),\quad \mathrm{for}\quad \alpha = 1 - \lambda/2. 
\end{equation}
The characteristic polynomial thus reads:
\begin{align}\label{Plambdaepsilon}
    P(\lambda,\epsilon) =  & (1+\epsilon-\lambda)\Big[ \mathrm{U}_{N-1}(\alpha) - \mathrm{U}_{N-2}(\alpha)\Big] \nonumber \\
    & -  \Big[ \mathrm{U}_{N-2}(\alpha) - \mathrm{U}_{N-3}(\alpha)\Big]~.
\end{align}
The next steps are as follows: rearrange the terms of (\ref{Plambdaepsilon}) in order to extract a term proportional to $(1-\epsilon)$, and a second term proportional to $\epsilon$. Both of these terms will contain sums of $\mathrm{U}_i(\alpha)-\mathrm{U}_{i-1}(\alpha)$, where $i=N-1,N-2$. We proceed to eliminate all instances of $\mathrm{U}_{N-3}(\alpha)$ via the recurrence relation (\href{https://dlmf.nist.gov/18.9}{DLMF Sec. 18.9} \cite{DLMF_NIST}):
\begin{equation}
    \mathrm{U}_{N-3}(\alpha) = 2\alpha\,\mathrm{U}_{N-2}(\alpha) - \mathrm{U}_{N-1}(\alpha).
\end{equation}
The Chebyshev polynomial of the first kind, $\mathrm{T}_k$, is then introduced using a combination of the two formulas (Eq. 3 of Sec. 10.11, p. 184 \cite{bateman_bateman}, \href{https://dlmf.nist.gov/3.11}{DLMF Sec. 3.11.7} \cite{DLMF_NIST}, \cite{chebyshev-det, chebyshev-det2}):
\begin{align}
\mathrm{T}_n(x) &= \mathrm{U}_n(x) - x \,\mathrm{U}_{n-1}(x), \\ \nonumber
\mathrm{T}_{n+1}(x) &= 2x \,\mathrm{T}_{n}(x)-T_{n-1}(x). 
\end{align}
In particular, in the previous expression for $P(\lambda,\epsilon)$ the following substitution is made:
\begin{align}
&(2\alpha-2)\mathrm{U}_{N-1}(\alpha) + \Big\{\mathrm{U}_{N-1}(\alpha)-\mathrm{U}_{N-2}(\alpha)\Big\} = \frac{1}{\beta}\mathrm{T}_{2N+1}(\beta),
\end{align}
where $\beta = \sqrt{1-\lambda/4}$. This yields the characteristic polynomial in \eqref{eq:char-poly}, which in the current special case reads:
\begin{equation}\label{eq:char-poly-special}
    P =  \Bigg[(2\alpha-2) \mathrm{U}_{N-1}(\alpha)\Bigg] (1-\epsilon)
    +\Bigg[ \beta^{-1} \mathrm{T}_{2N+1}(\beta) \Bigg] \epsilon~.
\end{equation}
The same steps can be followed for the characteristic polynomial of $- \Delta(\epsilon,0)$, which yields:

\begin{align}\nonumber
P & = (1-\epsilon)^2 \Big[(2\alpha - 2) \mathrm{U}_{N-1}(\alpha)\Big]  + \epsilon^2 \Big[ \mathrm{U}_N(\alpha)  \Big] \\ \nonumber
& + 2(1-\epsilon)\epsilon \Big[\beta^{-1}\mathrm{T}_{2N+1}(\beta)\Big]~.
\end{align}


\section{Eigenvector derivation}\label{evec-derivation}
In this appendix we derive the eigenvectors of $-\Delta(\epsilon,0)$, i.e.  \eqref{eq:evecs}, reproduced here:
\begin{align}
\label{eq:evec2}
    [v(\lambda)]_m & =  \epsilon \Big[\mathrm{U}_{m-1}(\alpha)\Big] \nonumber \\
    & + (1-\epsilon)\Big[\beta^{-1}\mathrm{T}_{2m-1}(\beta)\Big]~.
\end{align}
In this expression $\lambda$ is a fixed eigenvalue determined as a root of the characteristic polynomial  (\ref{eq:char-poly}), and $\alpha = 1-\lambda/2$, $\beta = \sqrt{1-\lambda/4}$, as before. The derivation relies on the fact that the components $[v(\lambda)]_m$ satisfy the same iteration scheme as subdeterminants of $-\Delta(\epsilon,0)$.

More precisely, the recurrence relation for $[v(\lambda)]_m$ is found by unpacking the matrix eigenvalue problem, and is particularly simple since $-\Delta(\epsilon,0)$ is tridiagonal:
\begin{equation}\label{recurrence}
\begin{cases}
[v(\lambda)]_2 = (1+\epsilon-\lambda)[v(\lambda)]_1 \\
[v(\lambda)]_k = (2-\lambda)[v(\lambda)]_{k-1} - [v(\lambda)]_{k-2},\quad 2<k<N \\
(1+\epsilon-\lambda)[v(\lambda)]_N = [v(\lambda)]_{N-1}.
\end{cases}
\end{equation}
We may assume, without loss of generality, that $v_1 = 1$ since the state can be normalized arbitrarily without changing the eigenvalue equation. In addition to this, we may define:
\begin{equation}\label{boundaryrecurrence}
[v(\lambda)]_{N+1} \equiv 0 .
\end{equation}
The scheme (\ref{recurrence}), (\ref{boundaryrecurrence}) is in fact identical to the recurrence relation for subdeterminants of $(-\Delta(\epsilon)-\lambda)$. That is, the values:
\begin{equation}\label{subdeterminants}
\mathrm{det}_k\Big[-\Delta(\epsilon)-\lambda\Big],\quad k=0,1,\dots,N,
\end{equation}
corresponding to the determinant of the $k\times k$ matrix formed from the first $k$ rows/columns of $(-\Delta(\epsilon)-\lambda)$.
For any matrix $M$, we define $\mathrm{det}_0[M] = 0$, and since $\lambda$ is an eigenvalue, it follows that
$$\mathrm{det}_N \Big[-\Delta(\epsilon,0)-\lambda\Big] =0.$$ 
The recurrence relation for determinants of tridiagonal matrices \cite{thematrixcookbook} can then be used to show that the scheme (\ref{recurrence}) also holds for $\det_k[-\Delta(\epsilon) - \lambda]$; in fact,
$$[v(\lambda)]_{k} = \mathrm{det}_{k-1}\Big[-\Delta(\epsilon)-\lambda\Big],\quad k=2,\dots,N.$$
It can be shown that these subdeterminants (\ref{subdeterminants}) have already been derived in (\ref{eq:char-poly-special}), so we have derived the formula for the eigenvectors.


\end{document}